# Unfolding of the Spectrum for Chaotic and Mixed Systems

**Ashraf A. Abul-Magd and Adel Y. Abul-Magd**
Faculty of Engineering Science, Sinai University, El-Arish, Egypt

**Abstract**

Random Matrix Theory (RMT) is capable of making predictions for the spectral fluctuations of a physical system only after removing the influence of the level density by unfolding the spectra. When the level density is known, unfolding is done by using the integrated level density to transform the eigenvalues into dimensionless variables with unit mean spacing. When it is not known, as in most practical cases, one usually approximates the level staircase function by a polynomial. We here study the effect of unfolding procedure on the spectral fluctuation of two systems for which the level density is known asymptotically. The first is a time-reversal-invariant chaotic system, which is modeled in RMT by a Gaussian Orthogonal Ensemble (GOE). The second is the case of chaotic systems in which $m$ quantum numbers remain almost undistorted in the early stage of the stochastic transition. The Hamiltonian of a system may be represented by a block diagonal matrix with $m$ blocks of the same size, in which each block is a GOE. Unfolding is done once by using the asymptotic level densities for the eigenvalues of the $m$ blocks and once by representing the integrated level density in terms of polynomials of different orders. We find that the spacing distribution of the eigenvalues shows a little sensitivity to the unfolding method. On the other hand, the variance of level number $\Sigma^2(L)$ is sensitive to the choice of the unfolding function. Unfolding that utilizes low order polynomials enhances $\Sigma^2(L)$ relative to the theoretical value, while the use of high order polynomial reduces it. The optimal value of the order of the unfolding polynomial depends on the dimension of the corresponding ensemble.

## I. INTRODUCTION

Random matrix theory [1, 2] provides a framework to describe the statistical properties of spectra for quantum systems, whose classical counterpart are chaotic. It models the Hamiltonian of the system by an ensemble of $N$-dimensional random matrices, conditioned by general symmetry constraints. For example, a time-reversal-invariant quantum system is represented by a GOE of random matrices. RMT is also used for the integrable system by representing the Hamiltonian as a real diagonal random matrix whose eigenvalues are drawn at random from a Gaussian, leading to Poisson Orthogonal Ensemble (POE) fluctuations for these elements [3, 4]. Nevertheless, it is well known that not all the regular systems have a Poissonian NNS distribution [5-7]. The two-dimensional harmonic oscillator is a classical example [8].

In general RMT does not apply to the mixed systems for which the classical phase space has separate regions for regular and chaotic motion. Some models are introduced to apply RMT to mixed systems (see, e.g., [9, 10] and references therein).These models allow studying the transition from integrability to chaos. One of these models assumes that some of the quantum numbers of the system are approximately conserved. In this case the spectrum is composed by a superposition of independent subspectra and the Hamiltonian is represented by a block-diagonal matrix [11].

RMT aims to explain the correlation between energy levels independently of the mean level spacing. For this purpose, it is more common to ''unfold'' the spectrum by means of transformation [12] involving the cumulated level density so that the mean level spacing is equal to one. The unfolding of the spectrum must be implemented to get rid of the non-universal properties (level density) and concentrate on the fluctuations properties of the spectrum, which display universal properties. In most of practical applications, the exact form of the cumulated density level is unknown. Unfolding is usually done by parameterizing the numerically obtained level density in the terms of a smooth function, typically a polynomial. The choice of the order of the unfolding polynomial introduces ambiguities at the unfolding procedure. Unfolding can also be done by several other ways [13-15].

The aim of the present study is to quantify the randomness which may be present in spectral statistics due to the unfolding polynomial order. For this purpose, we discuss two special cases for which the spectral density is known. One of which is a simple GOE and the other has a composite spectrum of independent GOE sequences. We unfold the spectra of each model once by using the exact expression for the level density and once by approximating the density by polynomials of different degrees. We study the effect of different choice of the functional form of level density on spectral correlations. In particular, we quantify the short and long term correlations between levels by the nearest-neighbor spacing distribution $P(s)$ and the variance $\Sigma^2(L)$, respectively.

## II. UNFOLDING OF THE SPECTRUM

The main aim of RMT is to describe the fluctuations of the energy spectra. Before studying of the fluctuations, we must separate the local level fluctuations from overall energy dependence of the level separation. The level density of a standard random matrix ensemble is not directly related to the physical level density of the investigated systems. Nevertheless, it is needed for the proper unfolding of the spectral fluctuation measures. Unfolding is usually done by calculating the cumulative spectral function $I(E)$ of the observed or computed spectra, which is defined as the number of levels below or at the energy $E$. This function is frequently referred to staircase function. It may be separated into an average part $I_{\text{ave}}(E)$, whose derivative is the mean level density, and a fluctuating part $I_{\text{fluc}}(E)$. $I_{\text{ave}}(E)$ is calculated for each matrix of the ensemble by running spectral average. Whenever the functional form of mean level density $\rho(E)$ is known, the mean cumulative spectral density can be obtained,

$$I_{\text{ave}}(E) = \int_{-\infty}^{E} dE' \, \rho(E'). \tag{1}$$

The unfolded spectrum is formed by introducing a dimensionless energy variable
$$\varepsilon_i = I_{\text{ave}}(E_i). \tag{2}$$
In this variable, the spectra possess mean level spacing unity everywhere. The spacing between two successive levels in the unfolded spectrum can be obtained by a Taylor expansion of $I_{\text{ave}}(E)$, which yields
$$\begin{aligned}\varepsilon_{i+1} - \varepsilon_i &= (E_{i+1} - E_i)\frac{dI_{\text{ave}}(E_i)}{dE_i} + \text{h. o. t.}\\ &= \frac{E_{i+1} - E_i}{D_i} + \text{h. o. t.}\end{aligned} \tag{3}$$
where $D_i = 1/\rho(E_i)$ is the mean level spacing in the vicinity of $E_i$. Equation (3) suggests that the direct relation between the original and unfolded spectra is valid when the higher order terms can be neglected, which is happens where the level density is slowly varying. This is usually true in the center of spectrum in most of the studied cases.

The mean level density can be estimated in some special cases. Spectra of billiards are unfolded in terms of level densities obtained from Weyl's semiclassical law [16], which relates the billiard area and circumference to the number of resonance frequencies below a given one. Nuclear spectra are often unfolded in terms of formulae for level density derived from the Fermi gas models [17].

### A. Asymptotic level densities

If the system is modeled by GOE, the mean level density for infinitely large matrices is given by Wigner's semi-circle law [18]:
$$\rho_{\text{GOE}}(N, E) = \begin{cases} \frac{2N}{\pi a^2}\sqrt{a^2 - E^2}, & \text{for } |E| \leq a \\ 0, & \text{for } |E| > a \end{cases}. \tag{4}$$
where $a$ is the radius of semi-circle and $N$ is the size of the matrices. The parameter $a$ is related to the standard deviation $\sigma$ of the off-diagonal elements of Hamiltonian matrix by $a = 2\sigma\sqrt{N}$. Integrating Eq. (4) yields
$$I_{\text{GOE}}(N, E) = N\left[\frac{1}{2} + \frac{E}{\pi a^2}\sqrt{a^2 - E^2} + \frac{1}{\pi}\arctan\left(\frac{E}{\sqrt{a^2 - E^2}}\right)\right], \text{ for } |E| \leq a. \tag{5}$$
Another class of systems with known level density is that of a chaotic system in which one or more quantum numbers may be considered as weakly conserved. For example, many atomic nuclei show apparent collective behavior. They are often modeled in terms of wave functions characterized by few quantum numbers as in the case of the interacting Boson model [19]. In this case the Hamiltonian of the system can be arranged in a block-diagonal form; each block represents coupling between states of the same value of the quantum number(s). The spectrum of system is composed of a superposition of independent sequences of sub-spectra; each one is composed of eigenvalues of one of the blocks of the matrix. If all the blocks have the same size, and if each can be modeled by a GOE, the mean level density of the total spectrum is given by
$$\rho_{m\text{GOE}}(N, E) = m\, \rho_{\text{GOE}}(\tfrac{N}{m}, E) \tag{6}$$

This model has been successfully applied in [11] to model systems in the initial phase of the transition from integrability to chaos.

Our purpose now is to find out how large should the matrices of the ensemble be in order that the asymptotic expressions (4) and (6) provide a reasonable approximation for the mean level density. We numerically generate an ensemble of $N$-dimensional GOE matrices, with $N$ varying from 40 to 1000 and compare their level densities represented in FIG. 1 by histograms. The predictions of Eq. (4) are represented by continuous lines.

The agreement between the numerical and asymptotic level densities is good for all values of $N$. In addition, we generate ensembles of matrices with same dimensions, composed of $m$ diagonal GOE blocks. These ensembles represent chaotic systems, in which $m$ of the eigenvalues are still conserved. In FIG. 1, we also show the numerically-generated level densities (before unfolding) of the composite spectra with $m = 2$, 4 and 8. These results are comparable to those obtained by using the asymptotic formulae of the mean level density for each block of the Hamiltonian which yields the level density in Eq. (6).

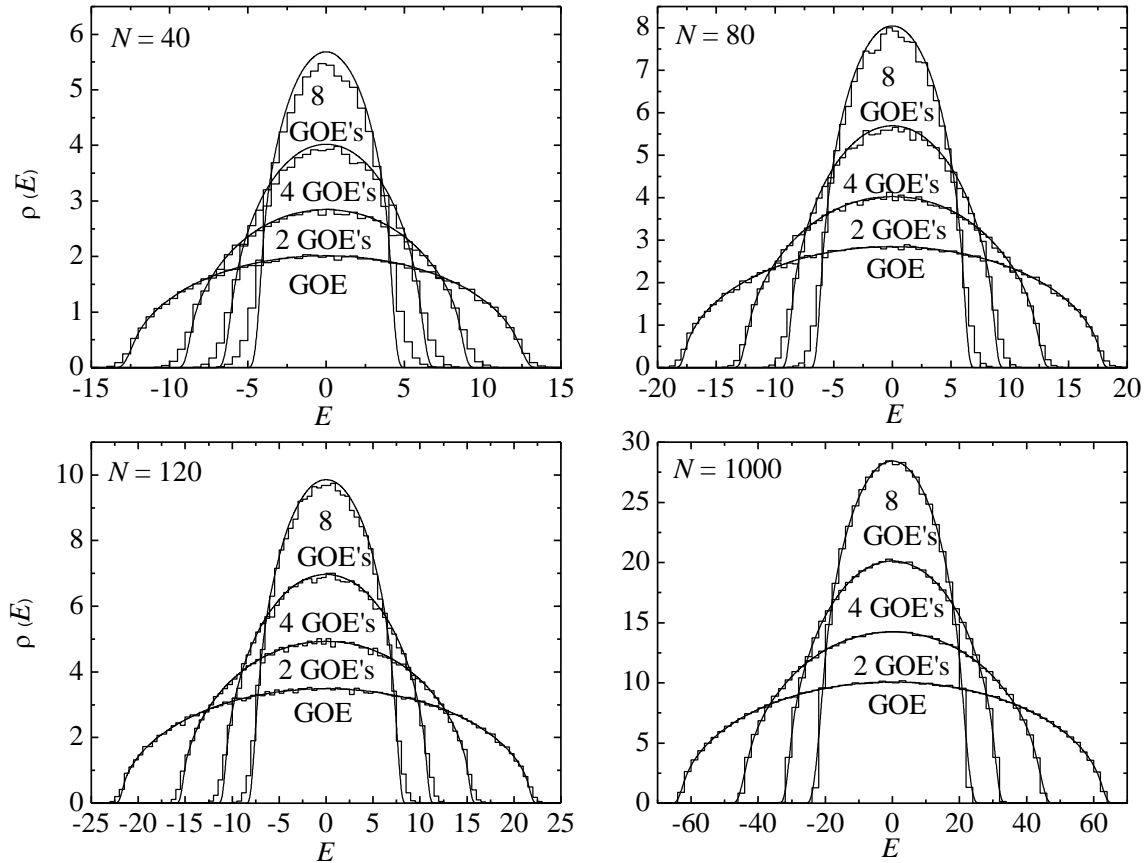

FIG.1. The level density $\rho(E)$ for both GOE and composite spectra of $m = 2$, 4 and 8 GOE's of $N$ matrices (histograms), compared to the asymptotic formulae (4) and (6) of the mean level density, which are obtained from Wigner's semi-circle law (solid line).

The figure shows that the degree of agreement deteriorates as $m$ increases for the smaller matrices. The asymptotic limit is reached practically at $N = 1000$ for all the four cases. To demonstrate this finding, we calculate the value of $\chi^2$ deviations [20] between the asymptotic formula and the numerical calculation. The $\chi^2$ is defined by

$$\chi^2 = \frac{4}{N}\sum_i \left(\frac{\rho_{\text{asymptotic}}(E_i) - \rho(E_i)}{\rho_{\text{asymptotic}}(E_i) + \rho(E_i)}\right)^2 \qquad (7)$$

where $\rho(E_i)$ are the numerical results for different $N$ matrices and $\rho_{\text{asymptotic}}(E_i)$ are the corresponding prediction of Eq. (6).

| $N$ | Number of ensemble members | Values of $\chi^2$ | | | |
|---|---|---|---|---|---|
| | | GOE | 2 GOE's | 4 GOE's | 8 GOE's |
| 40 | 2500 | 0.51 | 1.1 | 1.3 | 1.9 |
| 80 | 1250 | 0.39 | 0.58 | 0.95 | 1.3 |
| 120 | 850 | 0.28 | 0.46 | 0.80 | 1.05 |
| 1000 | 100 | 0.11 | 0.085 | 0.14 | 0.18 |

TABLE I. The values of $\chi^2$ for both GOE and composite spectra of $m = 2$, 4 and 8 GOE's, as a comparison between the asymptotic formulae in Eqs. (4, 6) and those calculated using $n$th degree polynomials in unfolding.

Table I show the value of $\chi^2$ as a measure of converging to the asymptotic formula. If we take the value $\chi^2 \leq 1$ as indicator for convergence, we conclude that the asymptotic expressions can be used for ensemble with $N$ larger than 120. Using them for ensembles of 1000×1000 matrices is certainly safe.

### B. Polynomial unfolding

When the exact form of the mean level density is unknown, one may smooth the staircase function by fitting it to a continuous function. Unless one knows the exact form of the level density, unfolding is not a unique procedure because there is no criterion whether the numerical estimated $I_{\text{ave}}(E)$ is close to the real one or not. Mostly, one fits $I_{\text{ave}}(E)$ to a polynomial of degree $n$. After extracting of the average part $I_{\text{ave}}(E)$, the observables are calculated for the unfolded spectrum of each matrix of the ensemble and subsequently, the results are averaged over the ensemble (the so-called spectral unfolding). This is done in the same spirit, as it is done in the nuclear data ensemble [17], where the spectrum of each nucleus is unfolded separately.

In early analysis of spectral fluctuations (e.g. in [21]), the number of levels in each member of the ensemble was small (between 5 and 20). The integrated level density (staircase) was fitted with a polynomial of third degree. Later, when spectra composed of large number of levels became available; many authors used polynomials of higher degrees in unfolding of the spectrum. For example, polynomials of order 15 were used in [15] and [22]. Soon it became clear that particularly the statistics that measure long-range level correlations were strongly dependent on the degree of the polynomials utilized in unfolding procedure. Increasing the order of polynomials utilized in unfolding does not always lead to the best results. Indeed, if one increases the order of polynomial to the extent that its curve passes by most of the points of the staircase function, one obtains an unfolded spectrum composed of nearly equally spaced levels, a picket fence! In fact, Flores *et al*. [22] have found that the spacing distribution and number variance of the $J = 0$ and $T = 0$ levels of the two-body random ensemble, which are supposedly chaotic, become in better agreement with the GOE predictions when the order of the unfolding polynomial is decreased from 15 to 7.

Our purpose now is to find out the optimum order of the unfolding polynomial for the analysis of both short and long range of the spectral fluctuations of chaotic and mixed systems. We have calculated the cumulative level densities for both GOE and composite spectra of $m = 2$, 4 and 8 GOE's of 100 1000×1000 matrices. The results are given by histograms in FIG. 2. We then fitted each histogram to polynomials of orders $n$ vary in from 3 to 15. The curves of the best-fit polynomials are also shown in FIG. 2 and the coefficients of these polynomials are given in Table II. We note that the agreement between the polynomials and the corresponding staircases is good particularly at the centers of the spectra. The quality of agreement depends on the order of the polynomials at the edge of the spectra, as shown by the inserts of FIG. 2. To avoid any possible ambiguities, we removed ten levels at each edge of the spectrum in the unfolding procedure, restricting the study of the spectral fluctuations to the center of the spectrum of each realization as it is commonly done.

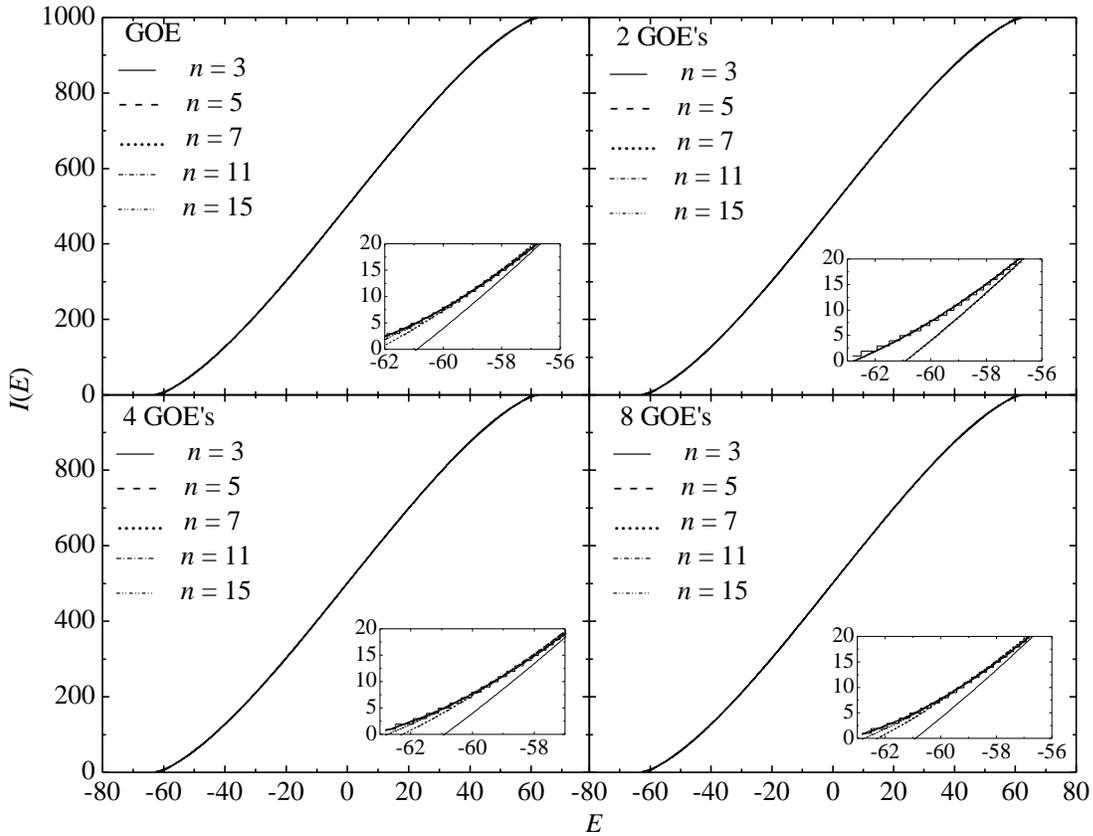

FIG. 2. The integrated level density (staircase) $I(E)$ for both GOE and composite spectra of $m = 2, 4$ and 8 GOE's (histograms), compared to fitting by polynomials of $n$ degrees. The inset shows a small section at the beginning of the spectrum.

|  |  | $a_0$ E+2 | $a_1$ E+1 | $a_2$ E-5 | $a_3$ E-4 | $a_4$ | $a_5$ | $a_6$ | $a_7$ | $a_8$ | $a_9$ | $a_{10}$ | $a_{11}$ | $a_{12}$ | $a_{13}$ | $a_{14}$ | $a_{15}$ |
|---|---|---|---|---|---|---|---|---|---|---|---|---|---|---|---|---|---|
| GOE | n=15 | 5.01 | 1.02 | 1.51 | -5.22 | -7.84E-17 | -1.09E-17 | 1.17E-19 | 1.31 | -8.35E-23 | -8.18E-24 | 3.09E-26 | 2.73E-27 | -5.70E-30 | -4.64E-31 | 4.14E-34 | 3.16E-35 |
|  | n=11 | 5.01 | 1.02 | 1.51 | -5.22 | 6.84E-19 | -2.01E-20 | -4.90E-22 | 1.30E-23 | 1.50E-25 | -3.55E-27 | -1.63E-29 | 3.40E-31 |  |  |  |  |
|  | n=7 | 5.01 | 1.02 | 1.51 | -5.22 | 4.00E-20 | -4.03E-22 | -3.20E-24 | -2.24E-26 |  |  |  |  |  |  |  |  |
|  | n=5 | 5.01 | 1.02 | 1.51 | -5.22 | 2.41E-20 | -7.60E-22 |  |  |  |  |  |  |  |  |  |  |
|  | n=3 | 5.01 | 1.02 | 1.51 | -5.22 |  |  |  |  |  |  |  |  |  |  |  |  |
| 2 GOE's | n=15 | 5.01 | 1.44 | -16.5 | -14.8 | -6.19E-16 | 3.78E-17 | 1.86E-18 | -9.03 | -2.67E-21 | 1.11E-22 | 1.98E-24 | -7.30E-26 | -7.26E-28 | 2.45E-29 | 1.05E-31 | -3.29E-33 |
|  | n=11 | 5.01 | 1.44 | -16.5 | -14.8 | 5.55E-18 | -1.62E-19 | -8.81E-21 | 2.24E-22 | 5.80E-24 | -1.42E-25 | -1.36E-27 | 3.33E-29 |  |  |  |  |
|  | n=7 | 5.01 | 1.44 | -16.5 | -14.8 | 9.81E-19 | -4.39E-20 | -5.70E-22 | 2.07E-22 |  |  |  |  |  |  |  |  |
|  | n=5 | 5.01 | 1.44 | -16.5 | -14.8 | -3.30E-19 | 1.22E-20 |  |  |  |  |  |  |  |  |  |  |
|  | n=3 | 5.01 | 1.44 | -16.5 | -14.8 |  |  |  |  |  |  |  |  |  |  |  |  |
| 4 GOE's | n=15 | 5.01 | 2.01 | -42.5 | -29.8 | 4.15E-7 | -1.21E-6 | -1.43E-17 | -1.32E-18 | 4.13 | 3.26E-21 | -6.14E-23 | -43.0 | 4.53E-26 | 2.89E-27 | -1.31E-29 | -7.74E-31 |
|  | n=11 | 5.01 | 2.01 | -42.5 | -29.8 | 4.15E-7 | -1.21E-6 | -8.36E-20 | 5.41E-21 | 1.11E-22 | -6.42E-24 | -5.10E-26 | 2.70E-27 |  |  |  |  |
|  | n=7 | 5.01 | 2.01 | -42.5 | -29.8 | 4.15E-7 | -1.21E-6 | -3.59E-21 | 1.59E-22 |  |  |  |  |  |  |  |  |
|  | n=5 | 5.01 | 2.01 | -42.5 | -29.8 | 4.15E-7 | -1.21E-6 |  |  |  |  |  |  |  |  |  |  |
|  | n=3 | 5.01 | 2.03 | -11.2 | -42.0 |  |  |  |  |  |  |  |  |  |  |  |  |
| 8 GOE's | n=15 | 5.00 | 2.87 | 27.5 | -118 | 9.45 | 1.70 | -1.04E-16 | -1.61E-17 | 55.8 | 7.82E-20 | -1.56E-21 | -2.03E-22 | 21.7 | 2.69 | -1.20E-27 | -1.42E-28 |
|  | n=11 | 5.00 | 2.87 | 27.5 | -118 | 3.98E-17 | -1.18E-17 | -2.30E-19 | 6.05 | 5.56E-22 | -1.32E-22 | -4.53E-25 | 1.02 |  |  |  |  |
|  | n=7 | 5.00 | 2.87 | 27.5 | -118 | -1.88E-18 | -5.11E-17 | 1.03E-20 | -2.72E-22 |  |  |  |  |  |  |  |  |
|  | n=5 | 5.00 | 2.87 | 27.5 | -118 | 3.16E-18 | -2.01E-17 |  |  |  |  |  |  |  |  |  |  |
|  | n=3 | 5.00 | 2.87 | 27.5 | -118 |  |  |  |  |  |  |  |  |  |  |  |  |

TABLE II. The coefficients of polynomials $f(E) = \sum_{k=0}^{n} a_k E^k$ used in fitting the cumulative level densities of GOE and composite spectra of $m = 2$, 4 and 8 GOE's.

It is difficult to judge the quality of the polynomial fit of the staircase function from FIG. 2. This is more clearly seen in FIG. 3, which shows the corresponding $\chi^2$ deviations of $I(E)$ in terms of $n$, defined as in Eq. (7). FIG. 3 shows that the value of $\chi^2$ does not depend strongly on the polynomial degree $n$ for the $P(s)$ (except maybe for the model with 4 GOEs).

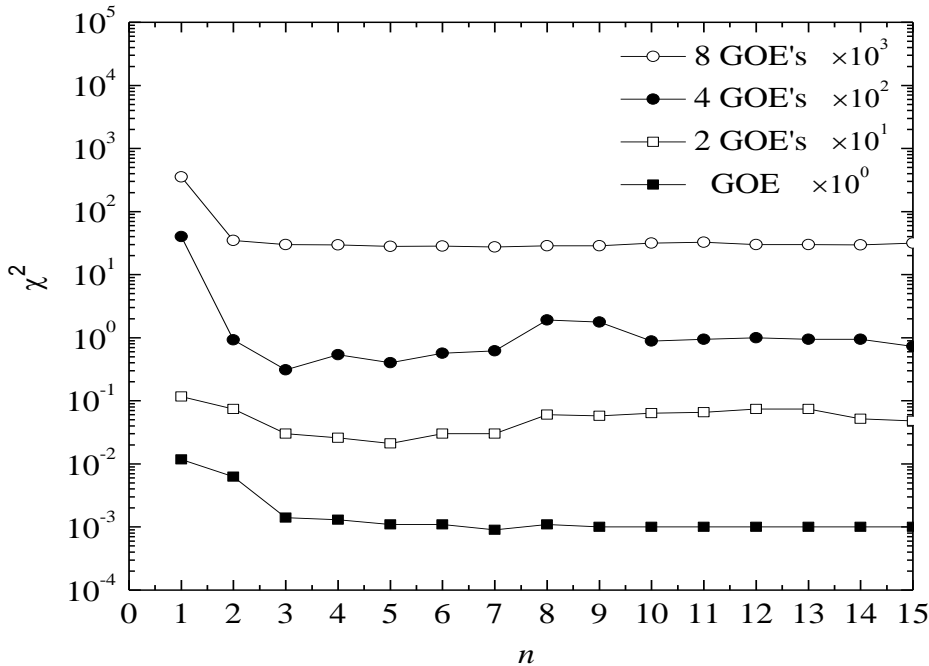

FIG. 3. The $\chi^2$ values qualifying the fitting of $I(E)$ by polynomials of $n$ degrees, for ensembles of 100 1000×1000 matrices of GOE and composite spectra of $m$ = 2, 4 and 8 GOE's.

### III. SENSITIVITY OF SPECTRAL FLUCTUATION TO UNFOLDING

#### A. Short range correlation

The nearest neighbor spacing distribution (NNSD) $P(s)$ is often used to study the short–range fluctuations in the spectrum. The matrix-element distribution is not directly useful to obtain the numerical results concerning energy-level statistics as NNSD. Let $E_1, E_2, \ldots, E_n$ be the position of successive levels at interval $\delta E$ ($E_1 < E_2 < \ldots < E_n$) and let $S_1, S_2,\ldots$ be their distances apart $S_i = E_{i+1} - E_i$. The average value of $S_i$ is the mean spacing $D$. We, also, define the relative spacing $s_i = S_i /D$. In practice, $s_i$ is taken as the difference between successive levels in the unfolded spectra (see Eq. (3) and the discussion following it). NNSD $P(s)$ is defined by condition that $P(s)\mathrm{d}s$ where the probability of any $s_i$ will have value between $s$ and $s$ + d$s$. RMT does not provide simple analytical expressions for NNSD. There are several elaborate approaches to evaluate this distribution. For example, Mehta [1] expresses the gap function as an infinite product involving eigenvalues of prolate spheroidal functions which are difficult to evaluate numerically. Another approach, which has been the subject of numerous investigations, is based on the relation to second-order nonlinear differential equations of the Painlevé type [22]. These approaches result in tabulated numerical values, series expansions and asymptotic expressions for NNSD. Analysis of discrete data is often done in terms of the spacing distribution of the two-dimensional GOE, which is given by [1]

$$P_{\text{GOE}}(s) = \frac{\pi}{2} s\, e^{-\frac{\pi}{4} s^2}, \qquad (8)$$

and known as Wigner's surmise. The two-dimensional GOE obviously ignores the long range correlations within the spectra of chaotic systems. However, it provides an accurate approximation for NNSD of large matrices. It has been checked against numerical calculations and rigorous bounds (see p. 171 of [1]).

Berry and Tabor [8] conjectured that the fluctuations of quantum systems, whose classical counterpart is completely integrable, are the same as those of an uncorrelated sequence of levels. Infinitely, the large independent-level sequence could be regarded as a Poisson random process. The NNSD for "generic" is given by

$$P(s) = \exp(-s). \qquad (9)$$

Berry and Tabor also pointed out that not all integrable system show a Poisson distribution. The two-dimensional harmonic oscillator is a classic example. The spacing distribution does not exist if the oscillator frequencies are commensurable. It is peaked at a non-zero value if the frequencies are incommensurable. This has been explained by Pandey and collaborators [24, 25], who use number theory to show that harmonic oscillators have a strong level repulsion and no fixed spacing distribution.

Integrable and chaotic systems are two extremes. Most of the physical systems fall in the wide class of mixed systems, where the motion at some parts of classical phase space is regular and is chaotic in other parts (see, e.g. [26] and references therein). Abul-Magd and Simbel [27] have studied systems in which the freedom degrees are divided into two noninteracting groups, one having chaotic dynamics and the other is regular. The Hamiltonian of such system is given as a sum of two terms, one for each group of freedom degrees. Thus, each eigenvalues of the total Hamiltonian is expressed as a sum of two eigenvalues corresponding to the two Hamiltonian terms. Then the spectrum is given by a superposition of the independent chaotic subspectra. Each subspectrum corresponds to one (or one set) of the quantum numbers of the Hamiltonian's regular component. In the case of equal-sized subspectra where each subspectrum satisfies the GOE statistics, the level density of the total spectrum is given by Eq. (5) and is given by

$$P(m,s) = \left[\text{Erfc}\left(\frac{\sqrt{\pi}}{2m} s\right)\right]^{m-2} e^{-\frac{\pi s^2}{4m^2}} \times \left[\left(1 - \frac{1}{m}\right) e^{-\frac{\pi s^2}{4m^2}} + \frac{\pi s}{2m^2} \text{Erfc}\left(\frac{\sqrt{\pi}}{2m} s\right)\right], (10)$$

where Erfc(.) is the complimentary error function. This formula may be used as a useful for systems with mixed regular-chaotic dynamics. It is easy to see that in the case of a single sequence, $P(1,s) = \frac{\pi}{2} s \exp(-\frac{\pi}{4} s^2)$, the Wigner surmise is recovered. On the other hand, $\lim_{m\to\infty} P(m,s) = e^{-s}$, as required for regular systems.

We would now like to test the accuracy of the unfolding procedure for short range correlation between levels. FIG. 4 shows by histograms NNSD's for ensembles of 100 real symmetric 1000×1000 block matrices with number of blocks $m = 1, 2, 4,$ and 8. For matrices of this size, we expect the asymptotic expressions (9) and (10) to be valid, as shown above. The spectra, which have been unfolded with the third-order polynomial, are shown by solid histograms. The dashed histograms are for unfolding carried out by the corresponding formulas, obtained from the semi-circle law. These results are comparable with the theoretical NNSD's (solid lines), which are given by Eqs. (8) and (10). The figure shows that unfolding with a third-order polynomial representation of the integrated level densities yields practically the same results as unfolding by the exact formulas of Eqs. (8) and (10).This can be clearly shown by calculating $\chi^2$ deviations between a theoretical GOE and the numerical calculation. The $\chi^2$ defined by

$$\chi^2 = \frac{4}{N}\sum_i \left(\frac{P_{\text{GOE}}(s_i) - P(s_i)}{P_{\text{GOE}}(s_i) + P(s_i)}\right)^2, \quad (11)$$

where $P(s_i)$ are the numerical results for NNSD at spacing's $s_i$ obtained by with either polynomial or exact unfolding. The values of $\chi^2$ for higher-order polynomial unfolding exhibit a little difference with those of the third-order polynomials.

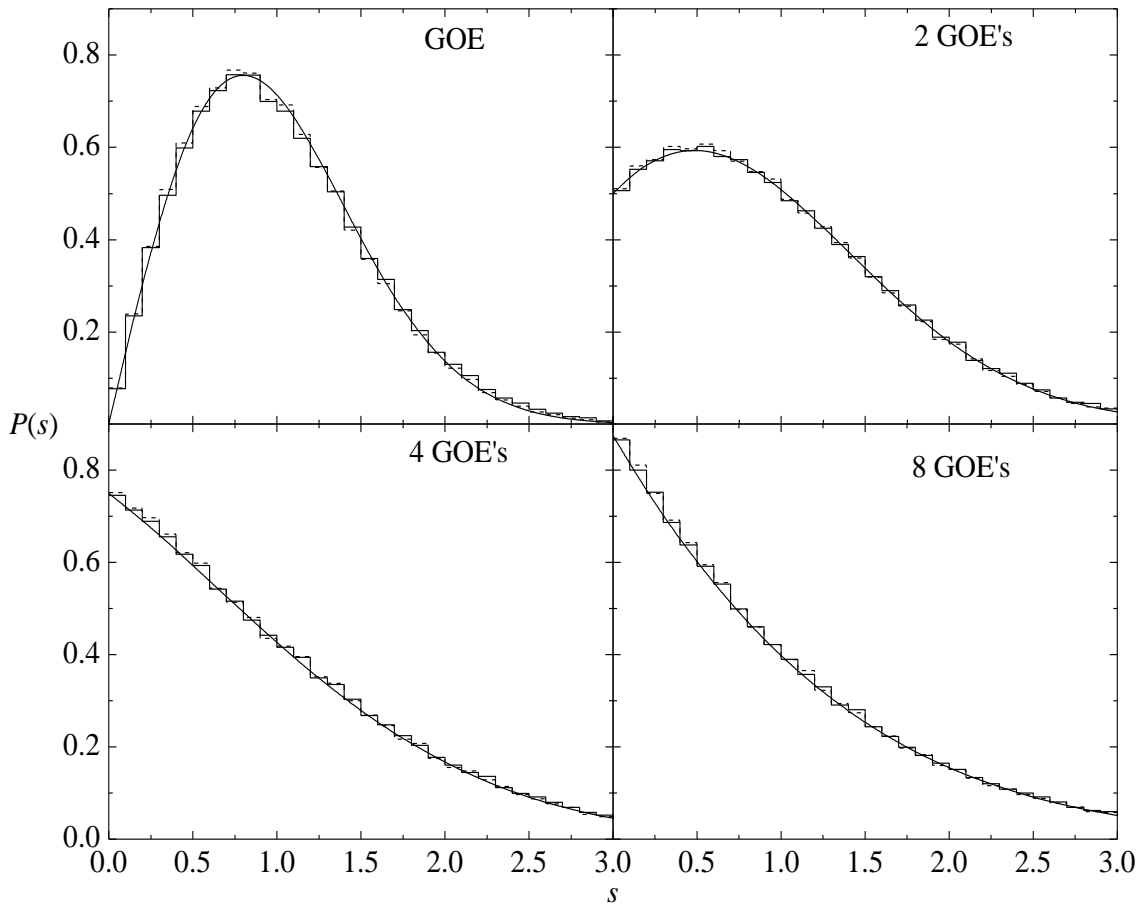

FIG. 4. NNSD for both GOE and composite spectra of $m = 2$, 4 and 8 GOE's of 100 1000×1000 matrices. The solid histograms are calculated using polynomial unfolding with $n = 3$ while the dashed ones are the numerical results of exact formula of semi-circular unfolding. The solid lines are calculated using Eq. (10).

## B. Long range correlations

The $\Sigma^2$ statistic, also called level-number variance (LNV), describes the long-range term correlation between levels. Specifically, for a given number $L$ of levels, of unfolded energy-levels in the interval $[\varepsilon, \varepsilon+L]$ is $n(L,\varepsilon) = I(\varepsilon+L) - I(\varepsilon)$, where $I(\varepsilon)$ is the integrated density of unfolded eigenvalues. LNV is defined by [28]

$$\Sigma^2(L,\varepsilon) = \langle [n(L,\varepsilon) - L]^2 \rangle_\varepsilon, \quad L > 0. \tag{12}$$

where $\langle . \rangle_\varepsilon$ denotes an average over $\varepsilon$. By definition, $\Sigma^2(L)$ measures LNV in an interval of length $L$ of the unfolded spectrum. In case of GOE, $\Sigma^2$ is given in Mehta's book [1] as

$$\Sigma^2_{\text{GOE}}(L) = \frac{2}{\pi^2}[\ln(2\pi L) - \text{Ci}(2\pi L)] + \gamma + 1 + \frac{1}{2}\text{Si}^2(\pi L) - \frac{\pi}{2}\text{Si}(\pi L) - \cos(2\pi L)$$
$$+ L\pi^2 \left(1 - \frac{2}{\pi}\text{Si}^2(2\pi L)\right) \tag{13}$$

where $\text{Si}(L)$ and $\text{Ci}(L)$ are the sine- and cosine-integral functions; respectively and $\gamma$ is an Euler's constant [29].

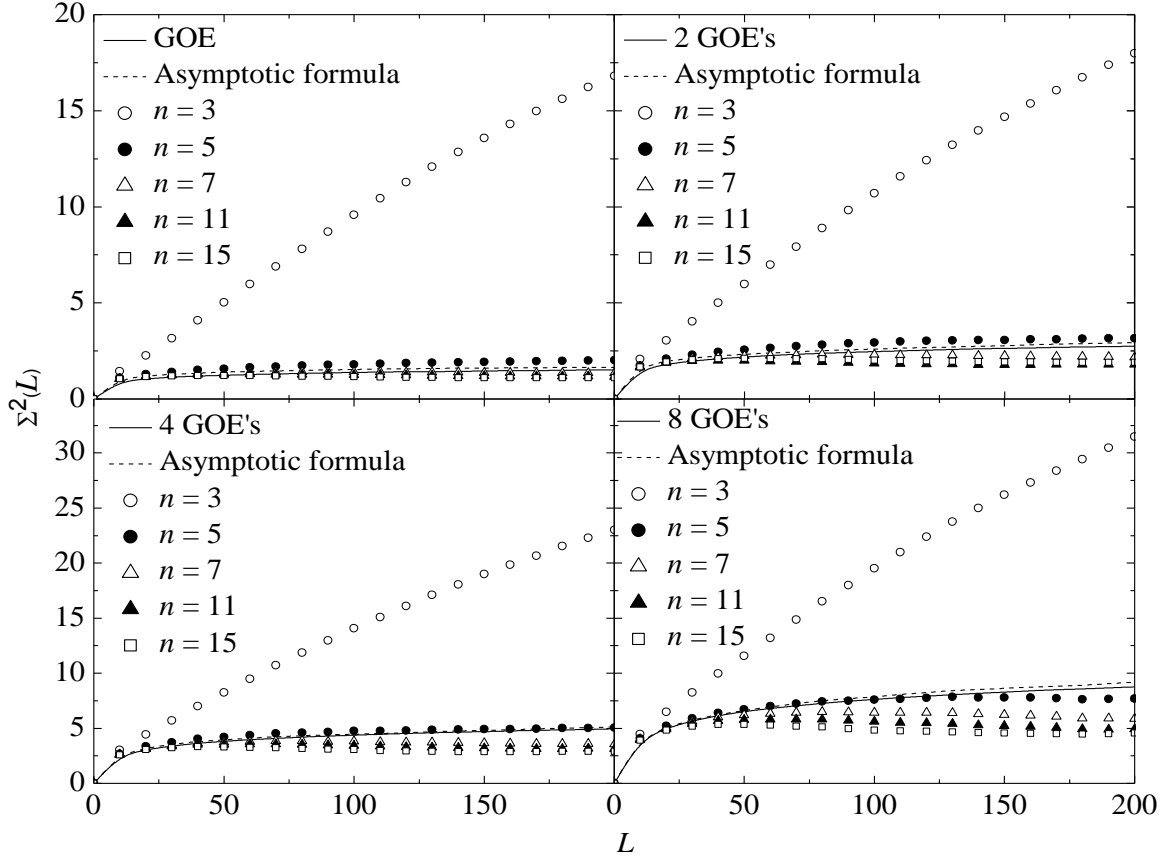

FIG. 5. LNV for both GOE and composite spectra of $m$ = 2, 4 and 8 GOE's of 100 1000×1000 matrices, calculated for $0 < L \leq 200$. The solid lines are calculated using Eq. (14), while the dashed ones are the numerical results of using the semi-circular formula in unfolding. Dots of different shapes are respectively the results of unfolding with polynomials of order $n$, shown on the figure.

The situation with LNV of composite spectra is not clear as in the case of NNSD [11]. The LNV of a composite spectrum composed of $m$ independent sequences of equal size is given by

$$\Sigma^2(m, L) = m\Sigma^2_{GOE}\left(\frac{L}{m}\right). \tag{14}$$

We now calculate numerically the value of $\Sigma^2(L)$ for the composite spectra of $m = 1$, 2, 4 and 8 GOE's. In order to demonstrate the influence of the unfolding of the spectrum, we compare the results of unfolding with polynomials of different degrees varying from $n = 3$ to 15, as well as the "exact" unfolding using the semi-circular formula in Eqs. (4) and (6). As FIG. 5 shows, unfolding in terms of the asymptotic formulas leads to good agreement with the numerical values of $\Sigma^2$ in the range of $0 < L \leq 200$. It also shows that the degree of the polynomial used in unfolding has a clear impact on LNV. When the degree of polynomial is 3, the calculated values of $\Sigma^2$ are higher than the numerical results especially at high values of $L$. In the degree range of 5-7, the agreement with the data is gradually improving. Further increase of the order of the unfolding polynomial produces lower values of $\Sigma^2$, tending to 0 which is the LNV of a picket-fence spectrum. FIG 6 shows the same results for $\Sigma^2(L)$ in the range of $0 < L \leq 20$, suggesting that the over-reduction of large values of $\Sigma^2$ does not occur when considering small values of $L$.

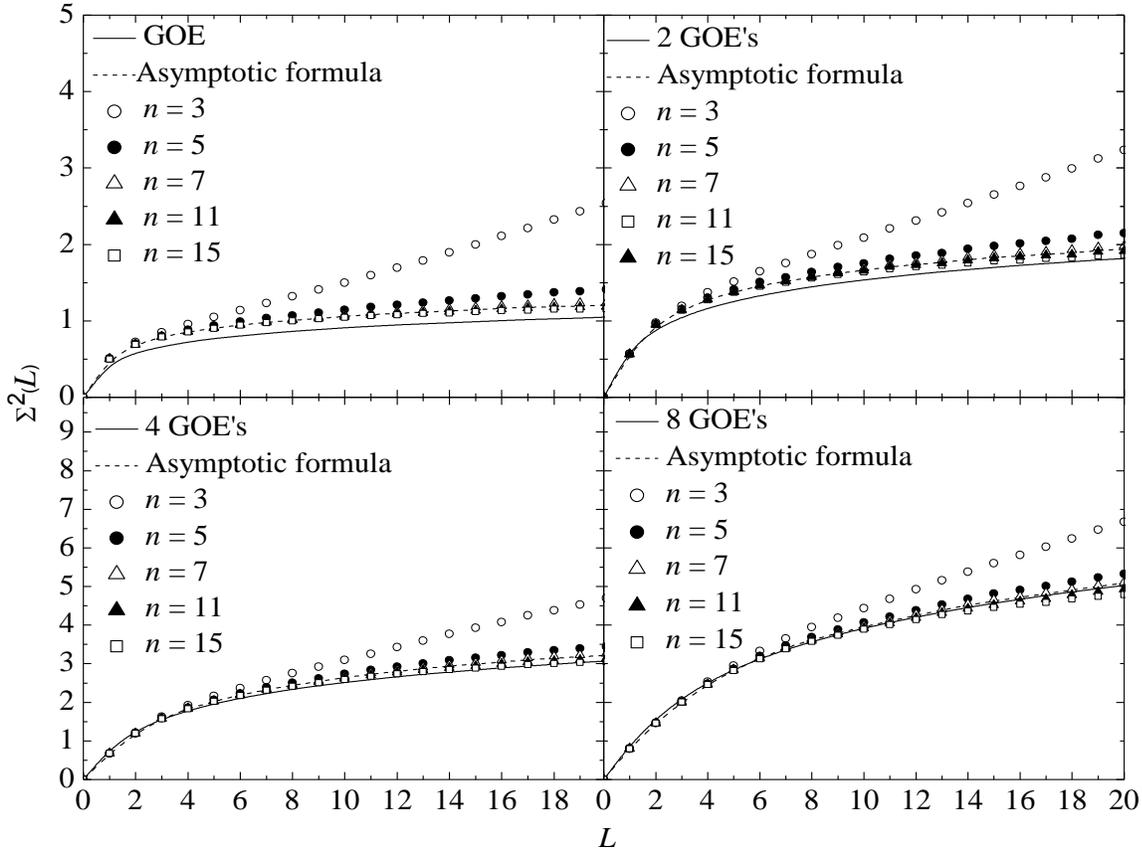

FIG. 6. LNV for both GOE and composite spectra of $m = 2$, 4 and 8 GOE's of 100 1000×1000 matrices, calculated for $0 < L \leq 20$. The solid lines are calculated using Eq. (14), while the dashed ones are the numerical results of using the semi-circular formula in unfolding. Dots of different shapes are respectively the results of unfolding with polynomials of order $n$, shown on the figure.

To clarify the behavior of $\Sigma^2$ in the range of $0 < L \leq 200$, FIG. 7 shows the value of $\chi^2$ deviations that are calculated by a formula analogous to Eq. (7). The values of $\chi^2$ show that the agreement between the numerical and asymptotic $\Sigma^2$ improves with increasing $n$ until the optimum value is reached. Further increasing the order of the unfolding polynomial reduces the agreement. As mentioned above, increasing $n$ yields better agreement between the integrated level density and the stair case function, which produces lower values of $\Sigma^2$, tending to 0 which is the LNV of a picket-fence spectrum. We also observe that the optimal order of the unfolding polynomial is rather sensitive to the size of the matrices of the ensemble. To show this, we repeated the previous calculation once for ensembles of 200 500×500 matrices and once for ensembles of 50 2000×2000 matrices for each of the studied models. FIG. 7 shows clearly that there is a more subtle dependence for $\Sigma^2(L)$. Interestingly, the minimum of $\chi^2$ for the 1000×1000 GOE is 7, which corresponds to the observation by Flores *et al*. [22] with essentially the same matrix size. Table III gives the orders of the unfolding polynomials that produce least values of $\chi^2$ in fitting LNV for the corresponding spectra to the asymptotic formulae in Eqs. (4) and (6). We may also conclude from FIG. 7 and Table III that systems with larger $m$, i.e. with more regular dynamics, prefer unfolding with lower order polynomials.

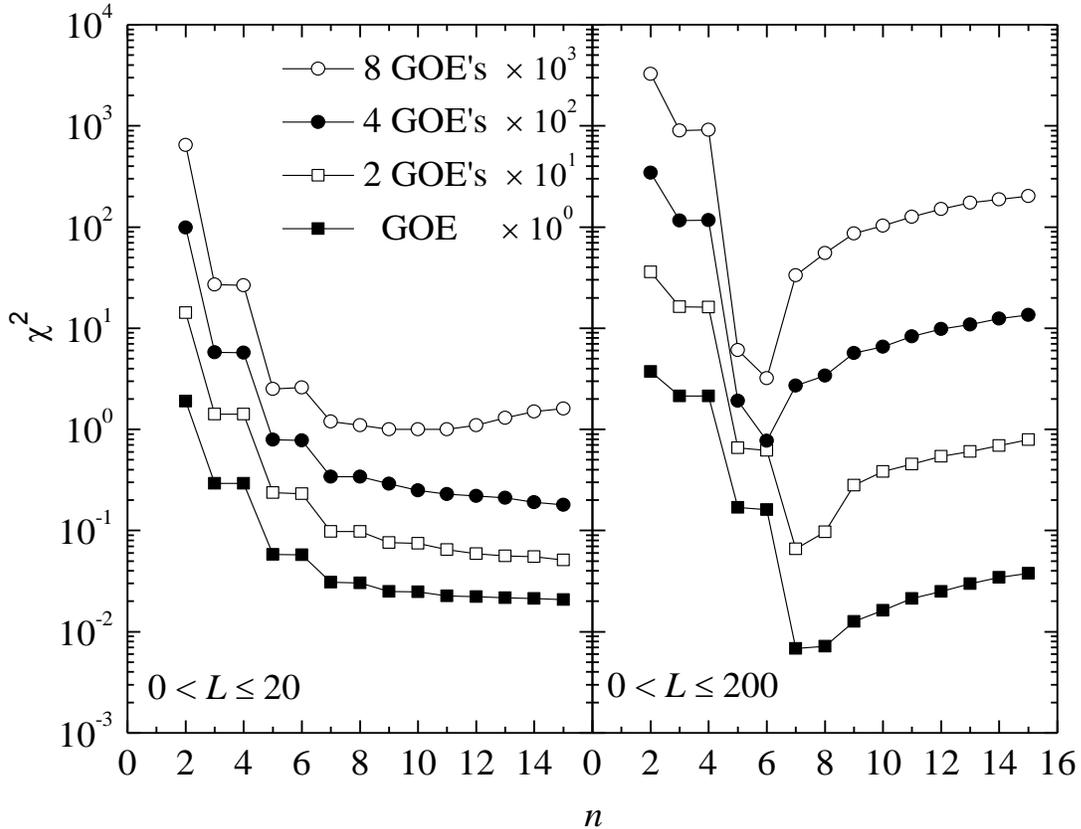

FIG. 7. The $\chi^2$ values qualifying the fitting of Eqs. (13) and (14) to LNV for ensembles of 100 1000×1000 matrices of GOE and composite spectra of $m = 2$, 4 and 8 GOE's for unfolded spectra that utilizes polynomials of order $n$.

| $N$ | Number of ensemble members | $n$ for least $\chi^2$ | | | |
|---|---|---|---|---|---|
| | | GOE | 2 GOE's | 4 GOE's | 8 GOE's |
| 500 | 200 | 6 | 5 | 5 | 5 |
| 1000 | 100 | 8 | 7 | 6 | 6 |
| 2000 | 50 | 11 | 9 | 7 | 8 |

TABLE III. Orders of the unfolding polynomials that produce least values of $\chi^2$ in fitting LNV calculated in the range of $0 < L \leq 200$ for GOE and composite spectra of $m = 2$, 4 and 8 GOE's to the asymptotic formulae in Eqs. (4, 6).

## IV. CONCLUSION

In this paper, we study the effect of choice of the functional form of the integrated level density, which are used in unfolding, on the spectral statistics of GOE and several GOE's. We show that changing the unfolding method has little effect on the short range correlations between levels as measured by NNSD. On the other hand, varying the degree of the polynomial representation of the mean level density used in unfolding has a strong influence on the long range correlation measured by LNV, $\Sigma^2$, and spectral rigidity $\Delta_3$. Our calculation shows that unfolding in terms of low order polynomials yields larger values of $\Sigma^2$, which gives an impression that the investigated system is more regular. On the other hand, if the polynomial degree is too big, then the unfolding procedure yields a spectrum, which looks more rigid than it should be. The optimal order of the unfolding polynomial slightly depends on the size of the spectrum under investigation. This suggests that much care should be taken in the unfolding of the spectrum while analyzing level number variance $\Sigma^2$ and spectral rigidity $\Delta_3$.

### Acknowledgement

The authors are grateful to S. Abulenein for his help in the numerical calculation.